\UseRawInputEncoding

\documentclass[aip, apl,10pt,twocolumn,superscriptaddress,10pt]{revtex4-1}
\usepackage{amssymb,amsmath,amsfonts,xcolor,graphicx}
\usepackage{times}
\usepackage{braket}
\usepackage{times}
\usepackage{float}
\usepackage{lipsum}

\begin{document}

\title{\Large Nonadiabatic geometric quantum computation with shortened path on superconducting circuits}

\author{Cheng-Yun Ding}
\affiliation{Guangdong Provincial Key Laboratory of Quantum Engineering and Quantum Materials,
and School of Physics and Telecommunication Engineering, South China Normal University, Guangzhou 510006, China}

\author{Yan Liang}
\affiliation{Guangdong Provincial Key Laboratory of Quantum Engineering and Quantum Materials,
and School of Physics and Telecommunication Engineering, South China Normal University, Guangzhou 510006, China}

\author{Kai-Zhi Yu} \email{Authors to whom correspondence should be addressed: kaizhiyu@163.com and zyxue83@163.com.}
\affiliation{Guangdong Provincial Key Laboratory of Quantum Engineering and Quantum Materials,
and School of Physics and Telecommunication Engineering, South China Normal University, Guangzhou 510006, China}

\author{Zheng-Yuan Xue}\email{Authors to whom correspondence should be addressed: kaizhiyu@163.com and zyxue83@163.com.}
\affiliation{Guangdong Provincial Key Laboratory of Quantum Engineering and Quantum Materials,
and School of Physics and Telecommunication Engineering, South China Normal University, Guangzhou 510006, China}

\affiliation{Guangdong-Hong Kong Joint Laboratory of Quantum Matter, and  Frontier Research Institute for Physics, South China Normal University, Guangzhou 510006, China}

\affiliation{Guangdong Provincial Key Laboratory of Quantum Science and Engineering, Southern University of Science and Technology, Shenzhen, Guangdong 518055, China}

\date{\today}

\begin{abstract}
Recently, nonadiabatic geometric quantum computation has been received much attention, due to its fast manipulation and intrinsic error-resilience characteristics. However, to obtain universal geometric quantum control, only limited and special evolution paths have been proposed, which usually requires longer gate-time and more operational steps, and thus leads to lower quality of the implemented quantum gates. Here, we present an effective scheme to find the shortest geometric path  under the conventional  conditions of geometric quantum computation, where high-fidelity and robust geometric gates can be realized by only single-loop evolution, and the gate performances are better than the corresponding dynamical ones. Furthermore, we can optimize the pulse shapes in our scheme to further shorten the gate-time, determined by how fast the path is travelled. In addition, we also present its physical implementation on superconducting circuits, consisting of capacitively coupled transmon qubits, where the  fidelities of geometric single- and two-qubit gates can be higher than  $99.95\%$ and  $99.80\%$ within the current state-of-the-art experimental technologies, respectively. These results indicate that our scheme is  promising for large-scale fault-tolerant quantum computation.
\end{abstract}

\maketitle

Different from classical computation, based on the nature of quantum coherence and entanglement, quantum computation can effectively solve several hard problems \cite{Shor, Grover} that cannot be solved by the classical one. To realize a  quantum computer, a set of universal quantum logic gates is the elementary building block \cite{Lloyd,Barenco}. However, the  decoherence and noise effects will inevitably  decrease the fidelity of the implemented quantum gates, which makes the computing invalid. Therefore, it's of central importance to explore  computing strategies  that are  intrinsically robust against decoherence, noise and local operational errors.

It is well-known that quantum  gates constructed by geometric phases, e.g., the adiabatic Berry phase \cite{Berry}, have strong robustness against certain local errors, as geometric phases  only depend on the global nature of evolution paths instead of the evolution details. This makes the  strategy  of constructing quantum  gates using   geometric phases very promising and it is termed as holonomic or geometric quantum computation (GQC) \cite{HQC, GQC} and has been experimentally demonstrated \cite{Jones,Wu,Huang}. However, due to the limitation of the  adiabatic process, the constructed geometric gates from Berry phase will suffer severely from the decoherence effect, resulting in unacceptable gate infidelity for typical solid-state systems. Besides, the adiabatic geometric phase has been generalized to the nonadiabatic case \cite{Aharonov}, which is more suitable in the application for constructing quantum gates. Therefore, GQC using the nonadiabatic geometric phase has been received extensive theoretical explorations \cite{Wang2001,Zhu12002, Zhu2003, Solinas1, Ota, Thomas, Zhao, Chen, Zhang1, Li2020, Chen1, lnji, Zhou, xu2020, wu2020, wang2020, saili, dingcy2021} and experimental demonstrations \cite{Leibfried,Du,Xu}.

To realize nonadiabatic GQC, the key step is to construct pure geometric gates, which means that accompanied dynamical phase during evolution process needs to be eliminated. The usual way before is to make the evolution follow some special paths on which the dynamical phase never exists \cite{Solinas1,Ota,Thomas}, especially for orange-slice-shaped loops \cite{Zhao,Chen,Zhang1,Zhou}.  However, these schemes  still have a considerable longer gate-time compared with corresponding dynamical gates, where decoherence will cause more gate infidelities. Moreover, to realize universal geometric gates, the trajectory there needs to mutate in the  South Pole  of the Bloch sphere, which complicates the operation process and causes more infidelities. Therefore, recent works aim to pursue shortening the needed time  for geometric  gates. For the non-cyclic geometric phase case \cite{Chen1, lnji}, using the time-optimal technique, the gate robustness of which is weaker than the cyclic one, this goal is achieved. But, for the cyclic case  \cite{Li2020, saili},  the shortened path still has mutation and/or the gate robustness is weak or untested. Besides,  there are other methods for non-adiabatic GQC with optimal control \cite{xu2020}, dynamical decoupling \cite{wu2020}, shortcuts to adiabaticity \cite{wang2020}, and path optimization  \cite{dingcy2021} to improve gate-robustness, but they need either longer gate-time and/or sudden mutative pulse control.  In addition, there is no comprehensive study of the shortest allowed path under certain conventional conditions.  Here, we propose a scheme for nonadiabatic GQC based on smooth circle paths, which are the shortest ones under the set conditions, and thus high-fidelity and robust geometric gates can be realized. Furthermore, pulse-optimization technique is also  applied to speedup the gate, as the gate-time is determined by how fast the path is travelled. Finally, we present the physical implementation of our protocol with superconducting  qubits and numerically shown the distinct gate performance. Therefore, these merits show that our scheme is a promising strategy in  realizing large-scale fault-tolerant quantum computation.

We now proceed to induce a pure geometric phase $\gamma_g$ by inverse design of a system Hamiltonian $\mathcal{H}(t)$ that satisfies the conditions of cyclic evolution and parallel transport. This phase is only related to polar angle $\alpha(t)$ and azimuth angle $\beta(t)$ of evolution states in the Bloch sphere, and thus to be geometric. Considering a set of auxiliary states as follows,
\begin{eqnarray}
\begin{split}
|\phi_{+}(t)\rangle &= \cos(\alpha(t)/2)|0\rangle+\sin(\alpha(t)/2)e^{i\beta(t)}|1\rangle ,    \\
|\phi_{-}(t)\rangle &= \sin(\alpha(t)/2)e^{-i\beta(t)}|0\rangle-\cos(\alpha(t)/2)|1\rangle,
\end{split}
\end{eqnarray}
where we require the evolution states $|\psi_{\pm}(t)\rangle=\exp[i\gamma_{\pm}(t)]|\phi_{\pm}(t)\rangle$ to satisfy the conditions of cyclic evolution  $|\psi_{\pm}(\tau)\rangle\!=\!\exp[i\gamma_{\pm}(\tau)]|\psi_{\pm}(0)\rangle$ and  parallel transport $\langle|\psi_{\pm}(t)\rangle|\mathcal{H}(t)|\psi_{\pm}(t)\rangle\!=\!0$, with $\gamma_{\pm}(\tau)$ being the accumulated total phases after an evolution time $\tau$. Then, the needed two-level Hamiltonian $\mathcal{H}(t)$ is calculated to be \cite{Li2020}
\begin{eqnarray}\label{Hamiltonian}
\mathcal{H}(t)&=&i\!\!\!\sum_{m,n=+,-}^{m\neq n}\langle\phi_{m}(t)|\dot{\phi}_n(t)\rangle|\phi_m(t)\rangle\langle\phi_n(t)| \nonumber \\
&=&\frac{\Delta(t)}{2}(|1\rangle\langle1|-|0\rangle\langle0|) +\left[\frac{\Omega(t)}{2}|1\rangle\langle0|+\textrm{H.c.}\right],
\end{eqnarray}
where the $\Delta(t)$ and $\Omega(t)$ can describe the detuning and Rabi frequency of driving microwave field, respectively, as
\begin{eqnarray} \label{para}
\Delta(t)&=&-\dot{\beta}(t)\sin^{2}\alpha(t), \notag\\
\Omega(t)&=& e^{i\beta(t)}[i\dot{\alpha}(t)-\dot{\beta}(t)\sin\alpha(t)\cos\alpha(t)].
\end{eqnarray}
The initial states $|\psi_{\pm}(0)\rangle=|\phi_{\pm}(0)\rangle$, after a period of cyclic evolution $\tau$, driven by the Hamiltonian $\mathcal{H}(t)$, evolve to the corresponding final states of $e^{\mp i\gamma_g(\tau)}|\psi_{\pm}(0)\rangle$. Accordingly, the evolution operator is
\begin{eqnarray}\label{evolution}
U(\tau)&=&e^{-i\gamma_g(\tau)}|\psi_{+}(0)\rangle\langle\psi_{+}(0)| +e^{i\gamma_g(\tau)}|\psi_{-}(0)\rangle\langle\psi_{-}(0)|   \nonumber\\
&=&e^{-i\gamma_g(\tau) \,\textbf{n}\cdot\boldsymbol{\sigma}},
\end{eqnarray}
where $\textbf{n}=(\sin\alpha_0\cos\beta_0, \sin\alpha_0\sin\beta_0, \cos\alpha_0)$ is the unit direction vector with $\alpha_0\!\!=\alpha(0), \beta_0\!=\beta(0)$, $\boldsymbol{\sigma}\!=\!\!(\sigma_x,\sigma_y,\sigma_z)$ is the Pauli vector, and geometric phase $\gamma_g(\tau)=\frac{1}{2}\int^{\tau}_{0}[1-\cos\alpha(t)]\dot{\beta}(t)dt$. $U(\tau)$ represents a quantum rotation operator with the rotation axis being $\textbf{n}$ by an angle of $2\gamma_g(\tau)$. As $\textbf{n}$ and $\gamma_g(\tau)$ can be arbitrary,  universal single-qubit geometric gates can be obtained.

\begin{figure}[tbp]
\centering
\includegraphics[width=1\linewidth]{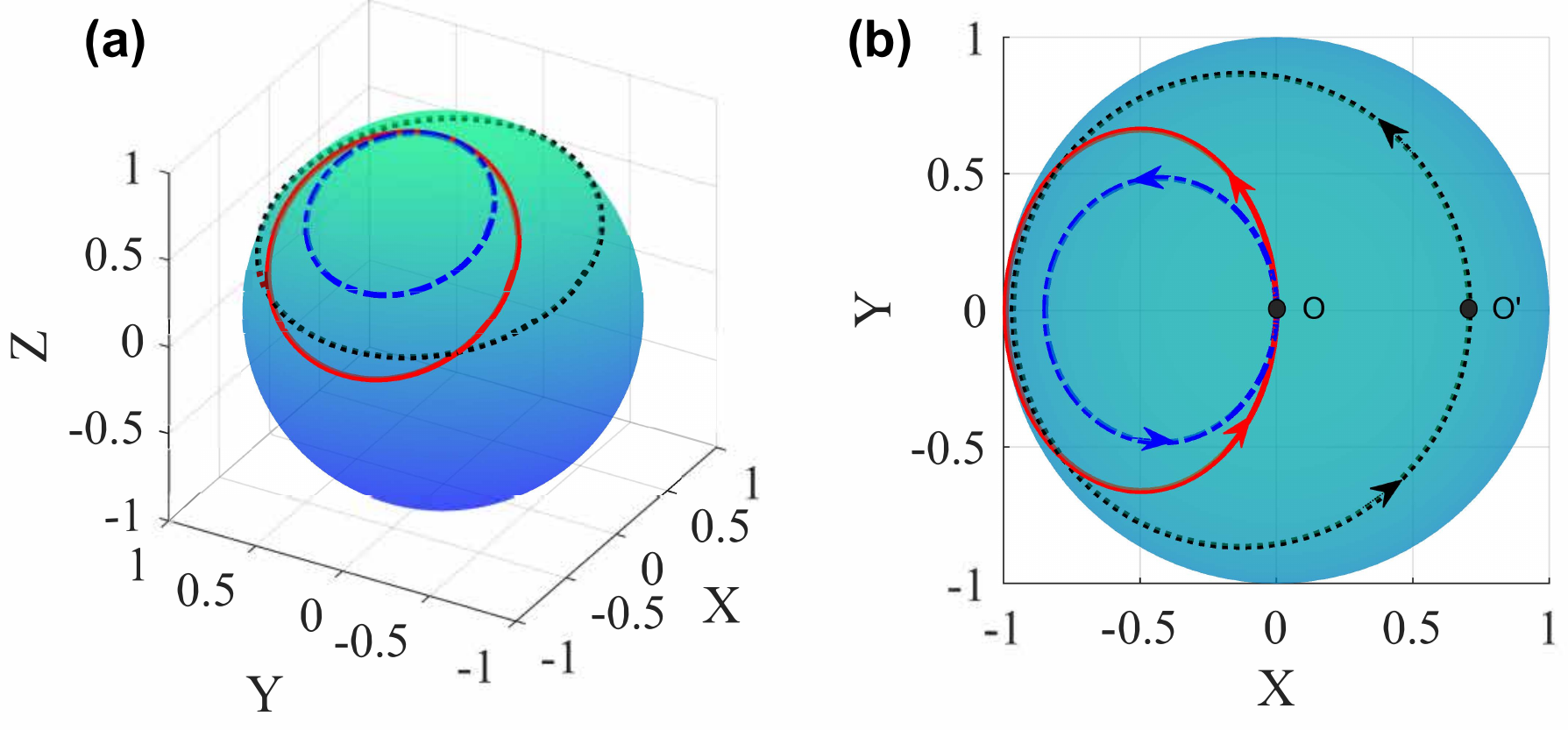}
\caption{Illustration of the shortest evolutionary trajectories for the  geometric Phase (solid-red line), $\pi/8$ (dash-dotted-blue line) and Hadamard (dotted-black line) gates, in Bloch sphere, respectively, for (a) $3$D view and (b) 2D view.}\label{Figure1}
\end{figure}

Next, we explore the equations of the shortest path for a set of general geometric gates. As shown above, for a given geometric phase $\gamma_g$ and starting point $(\alpha_0,\beta_0)$ , different parameters $\{\alpha(t),\beta(t)\}$ can be chosen, which corresponds to different evolution trajectories. In particular, a closed circle formed along a smooth trajectory is the shortest path among all the above trajectories, which generally represents a shortest gate-time for a given pulse shape. Specifically, according to Eq. (\ref{evolution}),  geometric  Phase, $\pi/8$ and Hadamard gates can be respectively constructed by setting following parameters,
\begin{eqnarray}
  \gamma_g&=&\frac{\pi}{4}, \alpha_0=0, \beta_0=\frac{\pi}{2}, \nonumber\\
  \gamma_g&=&\frac{\pi}{8}, \alpha_0=0, \beta_0=\frac{\pi}{2},  \\
  \gamma_g&=&\frac{\pi}{2}, \alpha_0=\frac{\pi}{4}, \beta_0=0. \nonumber
\end{eqnarray}
Note that the evolution of Phase and $\pi/8$ gates starts from the North Pole $O$. Besides, for a circle path, the following two conditions should be met: (i) the maximum value of polar angle $\alpha_m$ is unique and satisfies $\cos(\alpha_m/2)=1- \gamma_g/\pi$ for a given geometric phase and (ii) the change amount of azimuth angle $\beta(t)$ is $\pi$. Thus, in the case, a general parametric equation of the shortest path can be calculated as
\begin{equation}
\tan\frac{\alpha(t)}{2}=C\sin\left[\beta(t)-\frac{\pi}{2}\right],
\end{equation}
where $C\!=\!\!\sqrt{2\pi\gamma_g-\gamma^2_g}/(\pi-\gamma_g)$ is a constant. Meanwhile, the azimuth angle $\beta(t)$ can be set to take the form of
\begin{equation}\label{beta_ST}
\beta(t)=\frac{\pi}{2}+\pi\sin^{2}\left(\frac{\pi t}{2\tau}\right),
\end{equation}
for $t\in[0,\tau]$ with $\tau$ being the gate duration. Since the starting point of the Hadamard gate is   the point $O'(\pi/4,0)$, the above result can not be applied. But, we can still obtain an evolution equation that passes point $O'$ and the change of azimuth angle $\beta(t)$ is gradually increased from $0$ to $2\pi$. Then parameters' constrain will be
\begin{eqnarray}
&&2\sin\left(\frac{\pi}{12}\right)\sin\alpha(t)\cos\beta(t)\notag \\
&&\quad -2\cos\left(\frac{\pi}{12}\right)\cos\alpha(t)+1 = 0,   \\
&&\beta(t)=2\pi\sin^{2}\left(\frac{\pi t}{2\tau}\right). \label{beta_H}
\end{eqnarray}
The evolutionary trajectories for the  geometric gates, on the Bloch sphere, are plotted in Fig. \ref{Figure1}.
Then, according to  Eq. (\ref{para}), the Rabi frequency can be further calculated as
\begin{eqnarray}
\begin{split}
\Omega(t)=-\sqrt{\dot{\alpha}^2(t)+[\dot{\beta}(t)\sin\alpha(t)\cos\alpha(t)]^2}  \\
 \lefteqn{\times e^{i(\beta(t)-\zeta(t))}}\hspace{147pt} \\
=\Omega^{s}(t)e^{i(\beta(t)-\zeta(t)+\pi)},\qquad\qquad\qquad\;
\end{split}
\end{eqnarray}
with the pulse shape and time-dependent phase being
\begin{eqnarray}
\Omega^{s}(t)&=&\sqrt{\dot{\alpha}^2(t)+[\dot{\beta}(t)\sin\alpha(t)\cos\alpha(t)]^2},\nonumber \\
\zeta(t)&=&\arctan\left[\frac{\dot{\alpha}(t)}{\dot{\beta}(t)\sin\alpha(t)\cos\alpha(t)}\right].
\end{eqnarray}

\begin{figure}[tbp]
\centering
\includegraphics[width=0.95\linewidth]{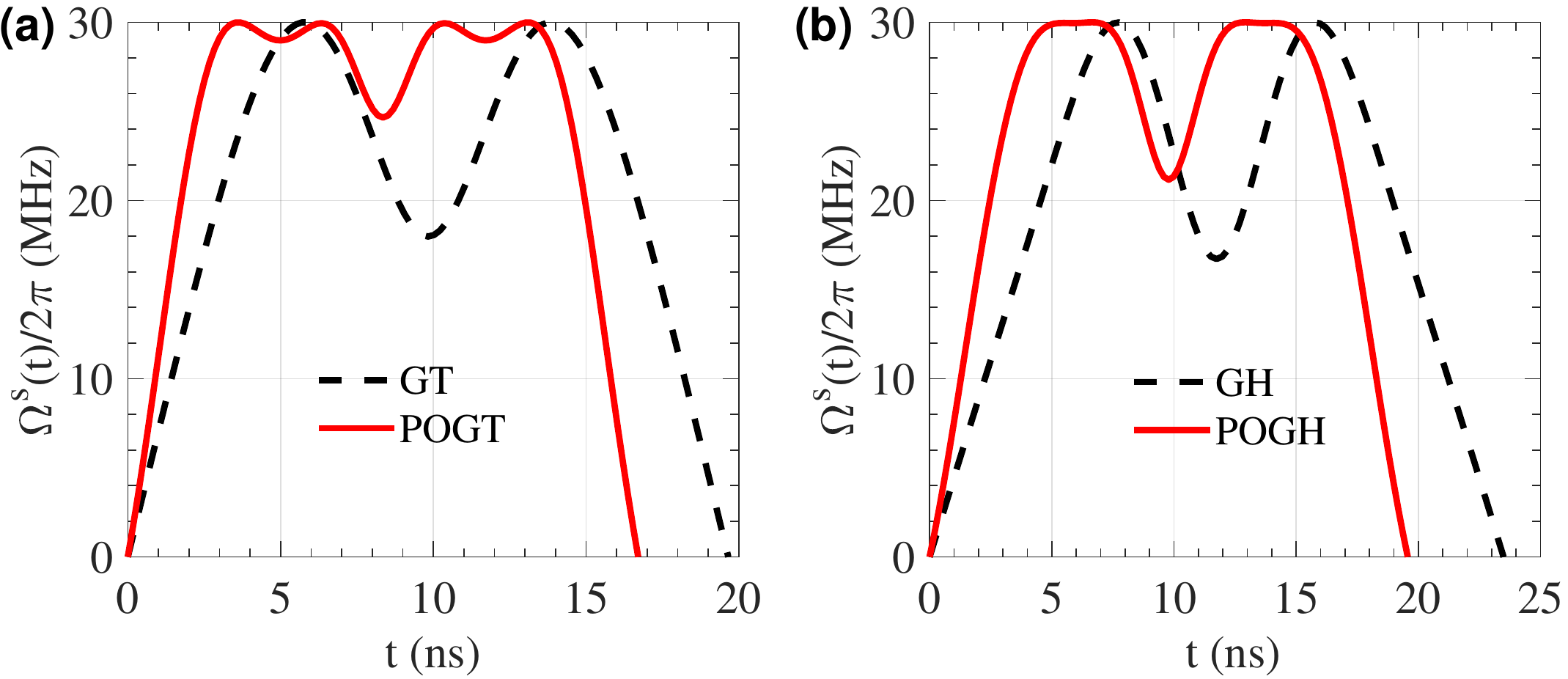}
\caption{Time dependence of the pulse strength $\Omega^{s}(t)$ for (a) geometric $\pi/8$ (GT) and (b) Hadamard (GH) gates, where solid-red (dashed-black) line denotes the pulse-optimized (PO) (pulse-unoptimized) case.}\label{Figure2}
\end{figure}

In addition, the simple forms of $\beta(t)$ in the Eqs. (\ref{beta_ST}) and (\ref{beta_H})  can also be in general forms as
\begin{eqnarray}\label{optimizedPulse}
\beta(t)&=&\frac{\pi}{2}+\pi\sin^{2}
\left(\frac{\pi t}{2\tau}\right)+\sum_{k=1}^{n}a_k\sin\left(\frac{2k\pi t}{\tau}\right),  \notag\\
\beta(t)&=&2\pi\sin^{2}\left(\frac{\pi t}{2\tau}\right)+\sum_{k=1}^{n}a_k\sin\left(\frac{2k\pi t}{\tau}\right),
\end{eqnarray}
where $a_k$ and $n$ (integral number) are free variables to modify the original shapes in order to  further shorten the needed gate durations. Here, $n=3$ is considered due to the technical limitation of actual pulse generator.  Since we only optimize the pulse with three additional term in Eq. (\ref{optimizedPulse}), other optimization methods \cite{Ruschhaupt2012,ding2021} with deliberate design may be used to further shorten the gate-time.  In Fig. \ref{Figure2}, we draw the unoptimized and optimized pulse shapes for geometric $\pi/8$ and Hadamard gates (Phase gate is similar to $\pi/8$ gate and is not shown hereafter). Meanwhile, the maximum pulse amplitudes are all set to be $\Omega^{s}_0=2\pi\times30$ MHz, and the optimal values of $a_k$ are listed in Table \ref{Table}. As shown in the Fig. \ref{Figure2},  the gate durations for geometric $\pi/8$ and Hadamard gates can be shortened from $19.66$ ns and $23.49$ ns to $16.71$ ns and $19.57$ ns, respectively.

\setlength{\tabcolsep}{7.5mm}{\begin{table}
\centering
\caption{Optimized values $a_k$ for geometric $\pi/8$ (GT) and Hadamard (GH) gates.}\label{Table}
\begin{tabular}{ccc}
\hline\hline\noalign{\smallskip}
Optimized values & GT&GH \\
\noalign{\smallskip}\hline\noalign{\smallskip}
$a_1$ & 0.007 & 0.095 \\
$a_2$ & 0.033 & 0.022 \\
$a_3$ & -0.024 & -0.046 \\
\noalign{\smallskip}\toprule
\end{tabular}
\end{table}}

\begin{figure}[tbp]
\centering
\includegraphics[width=0.95\linewidth]{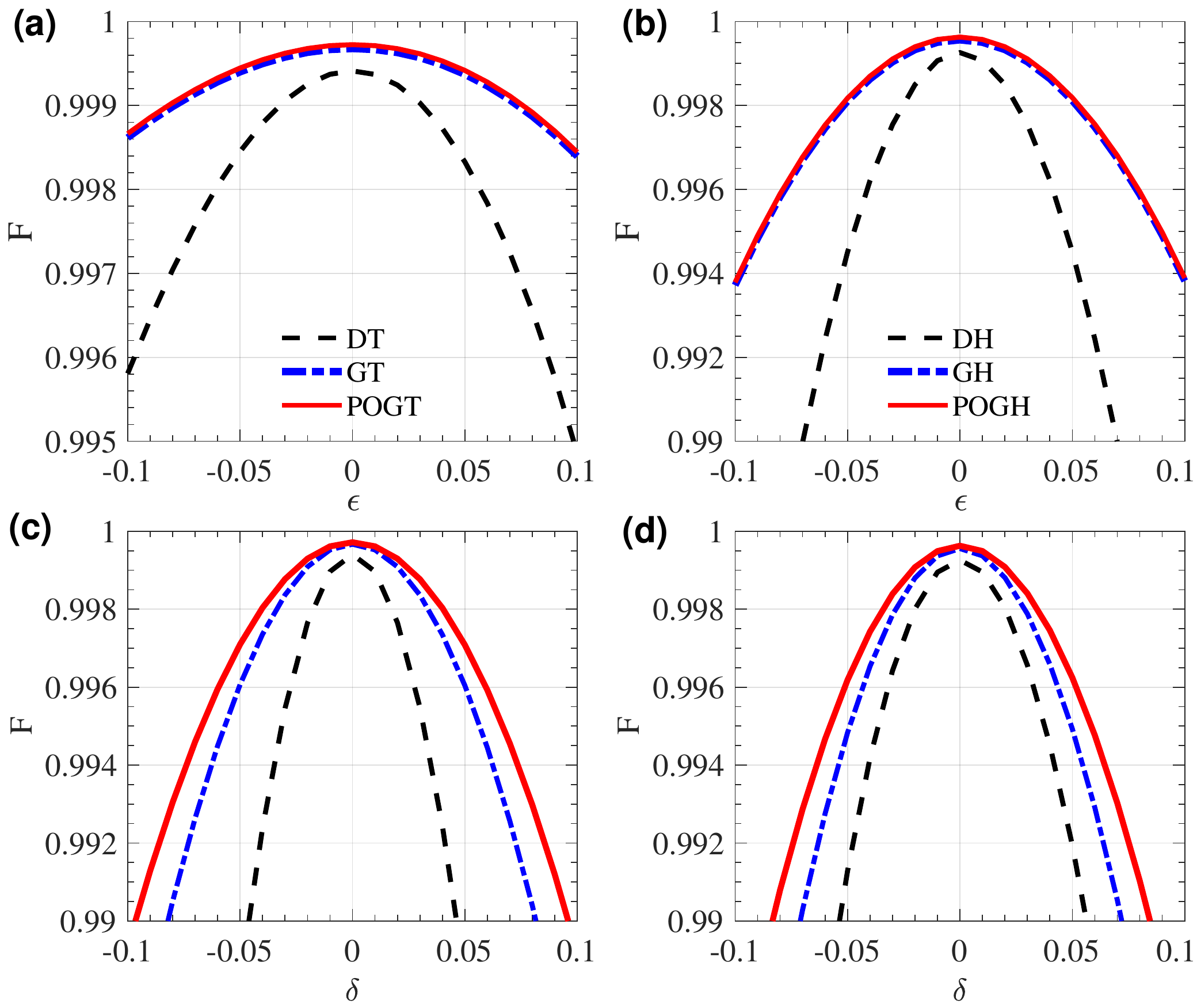}
\caption{Robustness comparison for (a), (c) geometric $\pi/8$ (GT) and (b), (d) Hadamard (GH) gates and the corresponding dynamical gates (DT, DH) against $\sigma_x$ and $\sigma_z$ errors, respectively, where solid-red lines denote the pulse-optimized (PO) geometric gates.}\label{Figure3}
\end{figure}

We now turn to consider the gate robustness of our scheme under decoherence.  Here, we mainly evaluate the robustness of geometric $\pi/8$ and Hadamard gates of our scheme against Rabi frequency and frequency detuning errors, i.e., $\sigma_x$ and $\sigma_z$ errors, and compare it with the corresponding dynamical gates. The dynamics of an open quantum system can be numerically simulated by using the Lindblad quantum master equation of \cite{Lindblad}
\begin{equation}\label{master}
\dot{\rho}=-i\left[\mathcal{H}'(t), \rho\right]+\frac{\Gamma}{2}\mathcal{L}(\sigma_{-})+\frac{\kappa}{2}\mathcal{L}(\sigma_{z}),
\end{equation}
where $\rho$ denotes density operator of the total quantum system, $\mathcal{H}'(t)=\mathcal{H}(t)+\frac{\epsilon}{2}\!\cdot\!\Omega(t)\sigma_x +\frac{\delta}{2}\!\cdot\!\Omega^{s}_0\sigma_z$ with $\epsilon,\delta\in[-0.1,0.1]$ being the error fractions of the Rabi frequency and detuning, respectively, $\mathcal{L}(\sigma)=2\sigma\rho\sigma^{\dag}-\sigma^{\dag}\sigma\rho-\rho\sigma^{\dag}\sigma$ is the Lindblad operator for operator $\sigma$, $\sigma_{-}=|0\rangle\langle1|$, $\sigma_z=|1\rangle\langle1|-|0\rangle\langle0|$, and $\Gamma$, $\kappa$ being decay and dephasing rates, respectively. To test the  robustness of the constructed geometric $\pi/8$ and Hadamard gates, we use a general  initial state of $|\nu^{1}_0\rangle=\cos\vartheta|0\rangle+\sin\vartheta|1\rangle$, after these gate operations, the ideal final states $|\nu^{1}_{\tau}\rangle$ will be $\cos\vartheta|0\rangle+\exp(i\pi/4)\sin\vartheta|1\rangle$ and $[(\cos\vartheta+\sin\vartheta)|0\rangle+(\cos\vartheta-\sin\vartheta)|1\rangle]/\sqrt{2}$, respectively. We define the gate fidelity as \cite{Poyatos} $F^{G}_1=\frac{1}{2\pi}\int^{2\pi}_{0}\langle\nu^{1}_{\tau}|\rho|\nu^{1}_{\tau}\rangle d\vartheta$ and the integration numerically performed for $1001$ initial states with $\vartheta$ being uniformly distributed over $[0, 2\pi]$. Assuming the decoherence rates \cite{Kjaergaard} to be $\Gamma=\kappa=2\pi\times3$ kHz,  the gate robustness is shown in Fig. \ref{Figure3}, which clearly indicates the performance improvement of our scheme comparing with the corresponding dynamical gates.

Finally, we propose the physical implementation of our scheme on superconducting quantum circuits consisted of capacitively coupled transmon qubits  \cite{Koch,You}. Considering that a transmon qubit is driven by an external microwave field, the system Hamiltonian can be written as
\begin{eqnarray}
\lefteqn{\mathcal{H}_1(t)=\frac{1}{2}\sum_{k=0}^{+\infty} \Big\{[(2k-1)\omega_0-k(k-1)\alpha]|k\rangle\langle k|}\hspace{245pt}  \nonumber \\
 \lefteqn{+\Omega^{s}(t)\sqrt{k}|k-1\rangle\langle k|e^{i[\int_0^t\omega(t')dt'-\phi(t)]}+\textrm{H.c} \Big\},}\hspace{218pt}
\end{eqnarray}
where $\omega_0$, $\alpha$ are the frequency and anharmonicity of the transmon qubit, $\Omega^{s}(t)$, $\phi(t)$ and $\omega(t)$ are the driving strength, phase and frequency of the microwave field, respectively. To analyze the Hamiltonian more clearly, we transfer it to the rotating framework with respect to frequency $\omega(t)$. It can be found that the obtained Hamiltonian is in the exact  same form as Eq. (\ref{Hamiltonian}) with $\omega_0-\omega(t)=\Delta(t)$ and $\phi(t)=\beta(t)-\zeta(t)+\pi$, in the qubit-subspace of $\{|0\rangle,|1\rangle\}$.

\begin{figure}
  \centering
  \includegraphics[width=0.95\linewidth]{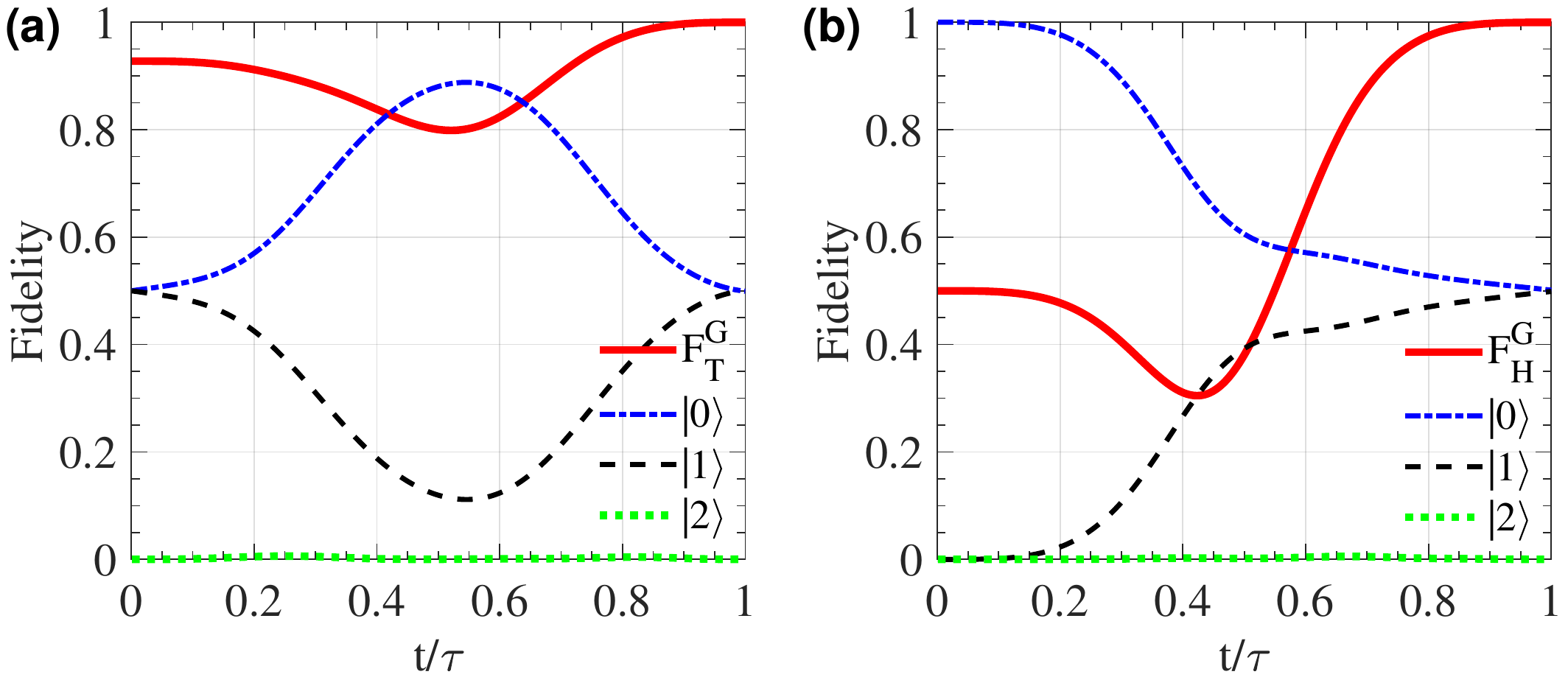}\\
  \caption{Dynamics of gate fidelity and state population for geometric (a) $\pi/8$ and (b) Hadamard gates, where the initial states are $(|0\rangle+|1\rangle)/\sqrt{2}$ and $|0\rangle$, respectively.}\label{Figure4}
\end{figure}

Note that, owing to the weak anharmonicity of transmon qubit, the driving field will induce not only energy transitions in the qubit subspace, but also cascade transitions to the non-computational subspace in a dispersive way, resulting in the information leaked out of the computational subspace. This leakage can be corrected by the ``derivative removal via adiabatic gate" technique \cite{Motzoi, Gambetta, Wang2018} such that a corrected pulse strength is $\Omega^{s}_{d}(t)=\Omega^{s}(t)-\{i\dot{\Omega}^{s}(t) +[\dot{\beta}(t)-\dot{\zeta}(t)+\Delta(t)]\Omega^{s}(t)\}/(2\alpha)$. Here, we consider $k = 0 ,1, 2$, that is, energy level $|2\rangle$ is regarded as the main leakage source, and use the master equation (\ref{master}) to evaluate the gate performance in practical superconducting system, with an available qubit anharmonicity \cite{Kjaergaard} of $\alpha=2\pi\times220$ MHz. For geometric $\pi/8$ and Hadamard gates, setting the parameters as in the general scheme case, the corrected gate fidelities can reach $99.96\%$ and $99.97\%$ respectively, as shown in Fig. \ref{Figure4}. At the same time, we set the initial states as $(|0\rangle+|1\rangle)/\sqrt{2}$ and $|0\rangle$ to see the change of its population over gate duration for the geometric $\pi/8$ and Hadamard gates, which is also shown in the Fig. \ref{Figure4}. Besides,  the used pulse shape $\Omega^{s}(t)$ here is the one before optimization, as the optimized pulse shape is closer to the square shape, which fails the leakage elimination technique. For quantum system with weak  leakage, the  optimized pulse will be more preferable.

Remarkably,  the implementation of nontrivial two-qubit geometric gates of our scheme can be achieved on superconducting quantum circuits consisted of two capacitively coupled transmon qubits, labeled by $Q_a$ and $Q_b$, respectively. Once their respective transition frequencies are set, their coupling strength $\textsl{g}$ is  generally fixed. To achieve controllable coupling \cite{Roth2017,Li2018,Cai,Chu}, we add an ac driving on transmon qubit $Q_b$ to adjust its qubit frequency $\omega_b$, so that $\omega_b(t)=\omega_b+\dot{f}(t)$ with $f(t)=\eta(t)\sin[\int_0^t\nu(t')dt'+\varphi(t)]$. Then, the interaction Hamiltonian of the two coupled transmon qubits can be calculated as
\begin{eqnarray}\label{two-qubit Hamiltonian}
\lefteqn{\mathcal{H}_{12}(t)=\textsl{g}\Big\{\big[|10\rangle_{ab}\langle01|e^{i\Delta t}+\sqrt{2}|11\rangle_{ab}\langle02|e^{i(\Delta+\alpha_b)t}+}\hspace{250pt} \nonumber \\
\lefteqn{\sqrt{2}|20\rangle_{ab}\langle11|e^{i(\Delta-\alpha_a)t}\big] e^{-i\eta(t)\sin[\int_0^t\nu(t')dt'+\varphi(t)]}\!+\!\textrm{H.c}\Big\}, }\hspace{250pt}
\end{eqnarray}
where $\alpha_a$ ($\alpha_b$) is the intrinsic anharmonicity of transmon qubit $Q_a$ ($Q_b$), and $\Delta=\omega_a-\omega_b$. Using the Jacobi-Anger identity $e^{i\beta \cos\alpha}=\sum^{+\infty}_{n=-\infty}i^{n}J_{n}(\beta)e^{i n\alpha}$ and
making a representation transformation with unitary operator $U(t)\!=\!\exp\big[\!-\!i\int_0^t\Delta'(t')dt'(|11\rangle_{ab}\langle11|-|02\rangle_{ab}\langle02|)/2\big]$ on Eq. (\ref{two-qubit Hamiltonian}), the transformed Hamiltonian will be
\begin{eqnarray}
\mathcal{H'}_{12}(t)&=&-\frac{\Delta'(t)}{2}|11\rangle_{ab}\langle11| +\frac{\Delta'(t)}{2}|02\rangle_{ab}\langle02| \nonumber \\
&+&\textsl{g}\Big\{\sum\limits^{+\infty}_{n=-\infty}i^{n}J_{n}[\eta(t)]e^{i [\int_0^t\nu(t')dt'+\varphi(t)+\frac{\pi}{2}]}\nonumber \\
&\times&\big[|10\rangle_{ab}\langle01|e^{i\Delta t}+\sqrt{2}|20\rangle_{ab}\langle11|e^{i(\Delta-\alpha_a)t}\nonumber \\
&+&\!\sqrt{2}|11\rangle_{ab}\langle02|e^{i\int_0^t[\Delta+\alpha_b+\Delta'(t')]dt'}\big] \!+\!\textrm{H.c}\!\Big\}.
\end{eqnarray}
When meeting the condition of $\nu(t)=\Delta'(t)+\alpha_b+\Delta$ and neglecting high-frequency oscillation terms, the above Hamiltonian reduces to
\begin{eqnarray}\label{Heff}
\mathcal{H'}_{eff}(t)=\frac{1}{2}\left(
                        \begin{array}{cc}
                          -\Delta'(t) & \textsl{g}'(t)e^{-i\varphi(t)} \\
                          \textsl{g}'(t)e^{i\varphi(t)} & \Delta'(t) \\
                        \end{array}
                      \right),
\end{eqnarray}
in a two-qubit subspace $\{|11\rangle_{ab},|02\rangle_{ab}\}$, with time-dependent $\textsl{g}'(t)=2\sqrt{2}\textsl{g}J_{1}[\eta(t)]$. It is obvious that Eq. (\ref{Heff}) has the same form as that of Eq. (\ref{Hamiltonian}), so state $|11\rangle_{ab}$ can be accumulated a pure geometric phase similarly to the single-qubit case, i.e., $|11\rangle_{ab}\!\!\rightarrow\!\! \exp(-i\gamma'_g)|11\rangle_{ab}$ in the finial time $\tau'$. Therefore, in the whole two-qubit subspace $\{|00\rangle_{ab},|01\rangle_{ab},|10\rangle_{ab},|11\rangle_{ab}\}$, we can obtain a two-qubit geometric control-phase gate from the Hamiltonian in Eq. (\ref{two-qubit Hamiltonian}), and the evolution operator is
\begin{equation}
U_2(\tau')=\text{diag}\left(
1, 1, 1, e^{-i\gamma'_g}
\right).
\end{equation}
Then, we test the performance of the two-qubit geometric gate by setting $\gamma'_{g}=\pi/4$ as a typical example. For a coupling strength of $\textsl{g}=2\pi\times10$ MHz, the optimized gate duration is $\tau'\simeq43.50$ ns corresponding to the maximal effective coupling strength $\textsl{g}'_{max}=2\pi\times15$ MHz, and $\eta(t)=J_{1}^{-1}[\textsl{g}'(t)/(2\sqrt{2}\textsl{g})]$ can be solved numerically. Other parameters of the transmon qubits are set within the current reachable range \cite{Kjaergaard}, i.e., $\Delta=2\pi\times500$ MHz, $\alpha_a=2\pi\times220$ MHz, $\alpha_b=2\pi\times200$ MHz and $\Gamma=\kappa=2\pi\times3$ kHz.  By these settings, the geometric control-phase gate fidelity can be numerically obtained to be $99.81\%$.

In summary, we have proposed a shortest-path-based scheme for universal nonadiabatic geometric quantum gates with higher fidelity and stronger robustness. Moreover, to pursuit shorter evolution time, the pulse-optimization technique can also be applied to the constructed geometric gates, especially for the nontrivial two-qubit geometric gates. Finally, fidelities for geometric single-and two-qubit gates can exceed $99.95\%$ and $99.80\%$ for the implementation on superconducting quantum circuits. Therefore, our scheme provides a promising way to achieve high-fidelity and robust GQC.

\bigskip
\noindent{\bf Authors' contributions}\\
C.-Y. D. and Y. L. contributed equally to this work.

\bigskip
\noindent{\bf Acknowledgements}\\
This work was supported by the Key-Area Research and Development Program of GuangDong Province (Grant No. 2018B030326001), the National Natural Science Foundation of China (Grant No. 11874156), Guangdong Provincial Key Laboratory of Quantum Science and Engineering (Grant No. 2019B121203002), and Science and Technology Program of Guangzhou (Grant No. 2019050001).

\bigskip
\noindent {\bf Data Availability Statement}\\
The data that support the findings of this study are available from the corresponding author upon reasonable request.

\end{document}